\documentclass[conference,10pt]{IEEEtran}
\usepackage{amsmath}
\usepackage{bm}
\usepackage{algorithm}
\usepackage{subfigure}
\usepackage{graphicx,epsfig,balance}
\graphicspath{{./figures/}} 
\usepackage{cite}
\usepackage{color}
\usepackage{multirow}
\usepackage{amsfonts,amssymb}
\usepackage{diagbox}
\usepackage[noend]{algpseudocode}
\usepackage{booktabs}
\usepackage{threeparttable}
\usepackage{diagbox}
\usepackage[noend]{algpseudocode}
\usepackage{algorithmicx,algorithm}
\usepackage{footnote}
\begin{document}

\title{Rogue Emitter Detection Using Hybrid Network of Denoising Autoencoder and Deep Metric Learning}

\author{Zeyang Yang$^\dag$, Xue Fu$^\dag$, Guan Gui$^\dag$, Yun Lin$^\star$, Haris Gacanin$^{\ddag}$, Hikmet Sari$^{\dag}$, and Fumiyuki Adachi$^{\dag\dag}$\\~\\	
$^\dag$College of Telecommunications and Information Engineering, NJUPT, Nanjing, China\\
$^\star$College of Information and Communication Engineering, Harbin Engineering University, Harbin, China\\
$^{\ddag}$Institute for Communication Technologies and Embedded Systems, RWTH Aachen University, Aachen, Germany\\
$^{\dag\dag}$International Research Institute of Disaster Science (IRIDeS), Tohoku University, Sendai, Japan\\
}
\maketitle

\begin{abstract}
Rogue emitter detection (RED) is a crucial technique to maintain secure internet of things applications. 
Existing deep learning-based RED methods have been proposed under the friendly environments. 
However, these methods perform unstable under low signal-to-noise ratio (SNR) scenarios. 
To address this problem, we propose a robust RED method, which is a hybrid network of denoising autoencoder and deep metric learning (DML). Specifically, denoising autoencoder is adopted to mitigate noise interference and then improve its robustness under low SNR while  
DML plays an important role to improve the feature discrimination.
Several typical experiments are conducted to evaluate the proposed RED method on an automatic dependent surveillance-Broadcast dataset and an IEEE 802.11 dataset and also to compare it with existing RED methods. Simulation results show that the proposed method achieves better RED performance and higher noise robustness with more discriminative semantic vectors than existing methods.
\end{abstract}


\IEEEpeerreviewmaketitle

\section{Introduction}
With the development of wireless communications, various internet of things (IoT) applications grow rapidly and play indispensable role for our daily life\cite{Chaudhary2019,Gui2020,Na2021}. However, the convergence of sensors, actuators, information, and communication technologies in IoT produces massive amounts of data that need to be sifted to facilitate reasonably accurate decision-making and control \cite{Chettri2019}. The openness of IoT makes it vulnerable to cybersecurity threats \cite{N.Wang2019,han2022network}. In particularly, identity spoofing attacks, where an adversary passively listens to the existing radio communications and then mimics the identity of legitimate devices to conduct malicious activities. 

Recently, radio frequency fingerprinting identification (RFFI) is a technique for identifying various RF devices by extracting inherent features from hardware defects in analog circuits \cite{Soltanieh2020}. These hardware imperfections appear during the manufacturing process. The most important merit of using physical imperfection as a signature for identification that it is hard to spoof the signature by using other wireless devices. RFFI has been used as an additional security layer for wireless devices to avoid spoofing or analog attacks \cite{Y.Xing2020,Meneghello2019}.

In recent years, deep learning-based specific emitter identification (SEI) methods have been proposed \cite{FuTCCN2021,Y.Wang2021,S.Chang2022,Pan2019,N.Yang2022}. These methods have their own advantages such as high identification accuracy, strong model generalization ability, and low model complexity. However, these methods are hard to apply in the field of rogue emitter detection (RED), which is important technique to solve the threat posed by the openness of IoT applications. 
Until know, some related works about RED have been investigated. Breunig \emph{et al.}\cite{Markus2000} and Liu \emph{et al.}\cite{Liu2008} respectively introduced LOF and IsolationForest, which are traditional RED methods based on machine learning. Bendale \emph{et al.}\cite{Bendale2016} used meta-recognition and replaced softmax with openmax, which managed the open space risk of deep networks while rejecting spoofed images. Akcay \emph{et al.}\cite{Akcay2018} introduced an encoder-decoder-encoder architecture for RED, and used the idea of generative adversarial training, which had good generalization ability for any RED task. Dong \emph{et al.} \cite{Y.Dong2021} presented SR2CNN, which could improve the feature discrimination by updating the semantic center vector. The details of these related works are shown in Table \ref{RELATEDWORK}. The above works mainly focused on improving the accuracy of RED at a specific high signal-to-noise ratio (SNR). However, existing RED methods are hard to work under the low SNR scenarios.
\begin{table*}
	\caption{Related works.}
	\label{RELATEDWORK}
	\centering
	\begin{tabular}{|c|c|c|c|c|c|}
		\hline
		\textbf{References}            & \textbf{Method} & \textbf{Data type}             & \textbf{RED} & \textbf{SNR/dB} & \textbf{Detection performance}                                                            \\
		\hline
		Y. Pan~\textit{et al.~}\cite{Pan2019} & DRN   & Singnal of radio emitters  & No       & $[10,24]$           & Acc: 56\%$\sim$95\%                                                                            \\
		\hline
		N. Yang~\textit{et al.~}\cite{N.Yang2022} & MAML  & ZigBee devices and 5 UAVs  & No           & $[0,30]$            & Acc: 30\%$\sim$100\%                                                                           \\
		\hline
		Bendale~\textit{et al.~}\cite{Bendale2016} & OpenMax & ImageNet (ILSVRC 2012 dataset) & Yes          & /               & ~F-measure: 0.595                                                                         \\
		\hline
		Akcay~\textit{et al.~}\cite{Akcay2018}   & GANomaly  & MNIST, CIFAR and X-ray & Yes          & /               & AUC: 0.666$\sim$0.882                                                                          \\
		\hline
		Y. Dong~\textit{et al.~}\cite{Y.Dong2021} & SR2CNN & RML2016.10a & Yes   & $\ge 16$  & \begin{tabular}[c]{@{}c@{}}True known rate: 0.959\\True unknown rate: 0.998\end{tabular}  \\
		\hline
	\end{tabular}
\end{table*}

To solve this problem, we propose a robust RED method by using a hybrid network of denoising autoencoder and deep metric learning (DML). Denoising autoencoder mitigates low SNR noise interference while the DML improves the feature discrimination so that the proposed method can further improve the RED performance in low SNR scenarios.
Furthermore, we introduce an objective function that consists of cross-entropy (CE) loss, mean squared error (MSE) loss and center (ML) loss, which allows the autoencoder to have better extraction performance of semantic features with high discrimination while saving feature space such that the proposed method has the potential to detect more rogue emitters.

\section{Signal Model and Problem Formulation}
\subsection{Signal Model}
In this paper, two typical datasets, i.e., ADS-B dataset and IEEE 802.11 dataset, are used to evaluate the performance of the proposed RED method. The receiver is used to receive ADS-B signals in a given airspace or IEEE 802.11 signals of the specific USRP transmitter. Assume that there are $K$ aircrafts or $K$ USRP transmitters transmitting signals, and the ADS-B signals from each aircraft or the IEEE 802.11 signals from each transmitter is received individually by the receiver. The received signal can be represented as:
\begin{align}
\label{align:signal model}
{r_{k}(t)=s_{k}(t) * h_{k}(t)+n_{k}(t), \quad k=1,2,\cdots, K}
\end{align}
where $r_{k}(t)$ is the received signal, $s_{k}(t)$ is the ADS-B signal by the aircraft or IEEE 802.11 signal transmitted by the transmitter, $h_{k}(t)$ is the channel impulse response between aircraft and receiver, $n_{k}(t)$ denotes the additive white Gaussian noise, and $*$ means the convolution operation.

\subsection{Problem Formulation}
Let $\mathcal X$ be the sample space and $\mathcal Y$ be the category space. One goal of the proposed RED method is to generate a mapping function $f \in {\mathcal F}:{\mathcal X} \rightarrow {\mathcal Y}$ , which can accurately predict the category of signals. ${\bf x} \in \mathcal X$ represents the signal. ${\rm y} \in \mathcal Y$ represents the true label of the signal. The mapping function $f$ can minimize the empirical error $\varepsilon_{e m}$, i.e.,

\subsubsection{Goal 1}
\begin{flalign}
&\label{align:signal model}
\min _{f \in {\mathcal F}} \varepsilon_{e m}=\min _{f \in {\mathcal F}} \mathbb{E}_{(\mathbf{x}, \mathrm{y}) \sim \mathcal{D}_{t}} \mathcal{L}_{CE}(f(\mathbf{x}), \mathrm{y})+\mathbb{E}_{(\mathbf{x}, \mathrm{y}) \sim \mathcal{D}_{\mathrm{t}}} \mathcal{L}_{\mathrm{red}}(\cdot)&
\end{flalign}
 where $\mathcal{D}_{t}$ represents the training dataset, $\mathcal{L}_{CE}$ represents the classification loss, and $\mathcal{L}_{\mathrm{red}}$ represents the regularization term which can prevent overfitting, improve the noise robustness, and extract more discriminable features for RED. MSE loss and ML loss are used as $\mathcal{L}_{\mathrm{red}}(\cdot)$ in this paper.

 According to the optimized mapping function, the semantic features of radio signals of legal emitters will be obtained and the semantic center features $\bm s_{k}$ can be calculated as:
 \begin{align}
\label{centerfeatures}
{{\bm s}_{k}=\frac{1}{N} \sum_{i=1}^{N} f_{en}({\bm x}_i), \text { if } {\bm x}_{i} \in\{ \text{class}~k}\},
\end{align}
 where $f_{en}$ denotes the mapping function of encoder, $k$ denotes the $k$-th category, and $N$ represents the number of samples of the $k$-th category. Our second goal is to detect whether the rogue emitters are legitimate by comparing the distance between semantic features of radio signals of unknown emitters and semantic center features, i.e.,

\subsubsection{Goal 2}
\begin{flalign}
\left\{\begin{array}{l}
\min \bm{d}(f_{en}(\bm{x}), \bm{S}) \leqslant \theta \Rightarrow \bm y \in \mathcal Y_{\text {in }} \\
\min \bm{d}(f_{en}(\bm{x}), \bm{S})>\theta \Rightarrow \bm y \in \mathcal Y_{\text {out }}
\end{array}\right.
\end{flalign}
 where $\bm S$ is the set of known semantic center features $\left\{{\bm s}_{k} \mid k=1,2,\cdots,K\right\}$, $\min \bm{d}(f_{en}(\bm{x}), \bm{S})$ represents the minimum distance between semantic features of radio signals of unknown emitters and the known semantic center features; $\theta$ represents the threshold for detection, $\mathcal Y_{\text {in }}$ represents the known category space, and $\mathcal Y_{\text {out }}$ represents the rogue category space.

\section{The Proposed RED Method}

\subsection{The Framework of Proposed RED Method}
\begin{figure}[htbp]
	\centering
	\includegraphics[width=3.5 in] {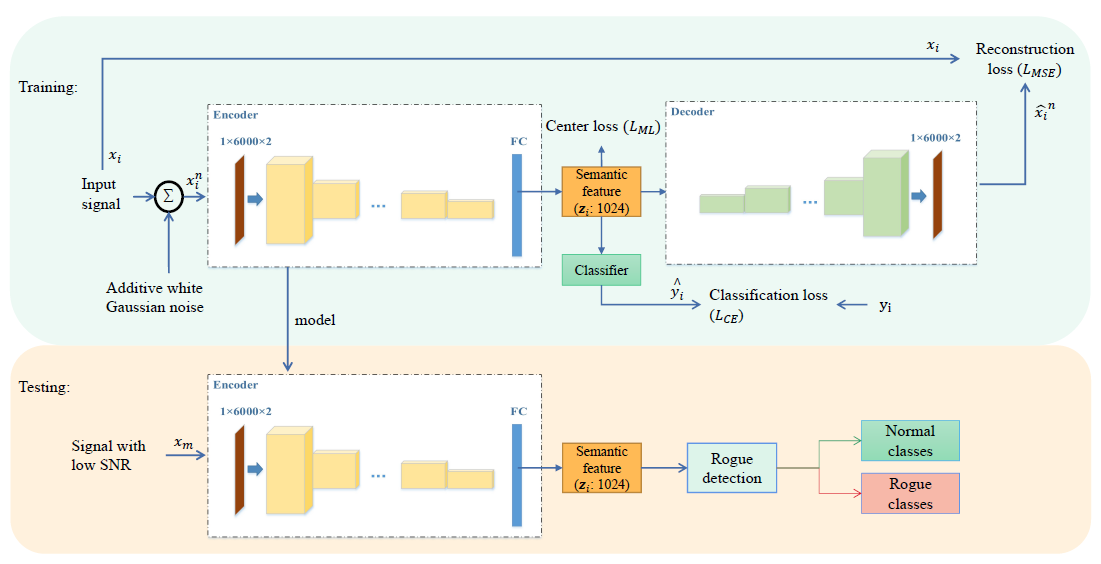}
	\caption{The framework of the proposed RED method.}
	\label{the framework of DMNet for RED}
\end{figure}

The proposed framework consists of an encoder, a decoder and a classifier, as shown in Fig. \ref{the framework of DMNet for RED}. The structure of each part of the network is shown in Table \ref{network}. Both encoder and decoder contain seven convolutional layers. Maxpool is a pooling operation, which reduces the feature dimension of the output of convolutional layers. BatchNorm is a batch normalization operation, which adjusts the distribution of the input values of each layer to a standard normal distribution and can speed up the training and convergence of the network. LazyLinear is a fully connected (FC) layer.

\begin{table}[htpb]
\caption{The network structure of proposed RED method.}\label{network}
\centering
\begin{tabular}{|c|c|c|}
\hline
\textbf{Net}                                   & \textbf{Layer}                                                                                   & \textbf{Number of layers} \\ \hline
\multicolumn{1}{|l|}{\multirow{2}{*}{Encoder}} & Input                                                                                            & $\times 1$                        \\ \cline{2-3}
\multicolumn{1}{|l|}{}                         & \begin{tabular}[c]{@{}c@{}}Conv (64, (10,1))+ ReLU \\ + BatchNorm + MaxPool ((4,1))\end{tabular} & $\times 7$                         \\ \hline
\multicolumn{1}{|l|}{\multirow{2}{*}{Decoder}} & \begin{tabular}[c]{@{}c@{}}ConvTranspose (64, (3,1))\\  + ReLU + BatchNorm\end{tabular}          & $\times 7$                         \\ \cline{2-3}
\multicolumn{1}{|l|}{}                         & Conv (1, (3,1)) + Sigmoid                                                                        & $\times 1$                         \\ \hline
\multirow{3}{*}{Classifier}                    & Flatten                                                                                          & $\times 1$                         \\ \cline{2-3}
                                               & LazyLinear (1024)                                                                                & $\times 1$                         \\ \cline{2-3}
                                               & LazyLinear (n classes)                                                                           & $\times 1$                         \\ \hline
\end{tabular}

\end{table}

We introduce an objective loss function using CE loss, MSE loss and ML loss. Specifically, we choose the CE loss to evaluate the classification loss, and it is the key part of objective loss function and it can be expressed as:
\begin{align}
\label{Classfication loss}
\mathcal L_{C E}=-{\mathbb E}(\bm y \log \left({\hat{\bm y}}\right)),
\end{align}
where $\bm y$ denotes the true label of training sample, $\hat{\bm y}$ denotes the predicted label of training sample.
To further obtain the model with strong robustness and extraction capability of discriminative features, the decoder is connected with encoder and two terms are used as regularization of objective function,
\begin{align}
\label{objective function}
\operatorname{min}\mathcal L=\operatorname{min}\{\lambda_{C E} \mathcal L_{C E}+\lambda_{M L} \mathcal L_{M L}+\lambda_{M S E} \mathcal L_{M S E}\},
\end{align}
where the $\mathcal L_{M L}$ is the ML loss; $\mathcal L_{M S E}$ is the MSE loss; $\lambda_{C E}$, $\lambda_{M S E}$, and $\lambda_{M L}$ are the weighting coefficients. With the regularization of ML loss, the model can obtain a set of network parameters suitable for mining discriminative semantic features. With the regularization of MSE loss, the model can obtain a set of network parameters suitable for improving the noise robustness. Two regularization terms will be introduced in details in the following two subsections. The fully training procedure for the proposed RED method is decribed in {\bf Algorithm \ref{alg:Training Procedure}}. After the training, the decoder and classifier will be discarded and the encoder is used for RED.

\subsection{Denoising Reconstruction for Strong Noise Robustness}
\begin{figure}[htbp]
  \centering
  \includegraphics[width=3.5 in]{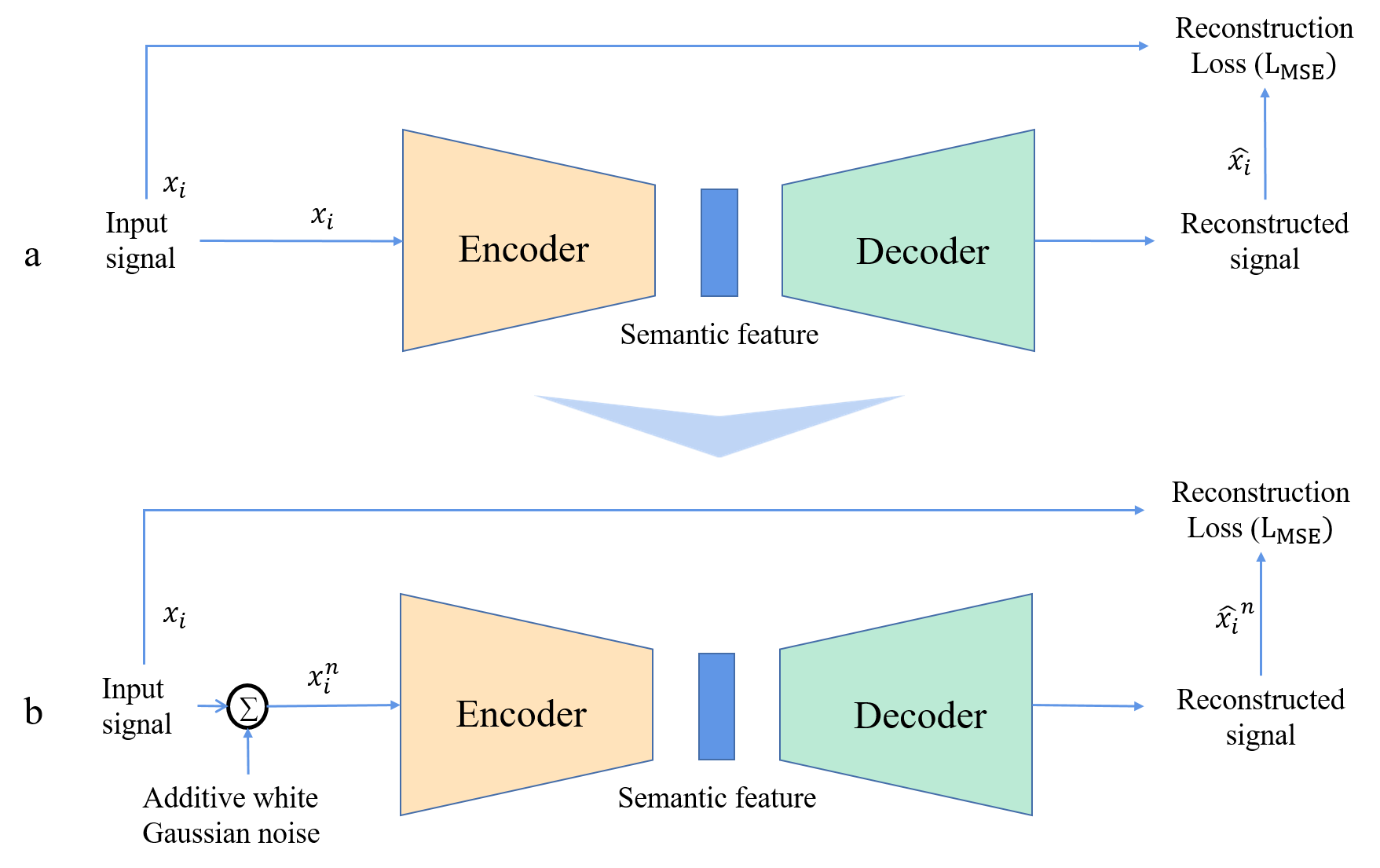}
  \caption{The structure of traditional autoencoder and denoising autoencoder of the proposed RED method.}
  \label{denosing}
\end{figure}

To improve the noise robustness, this paper adopts a denoising autoencoder architecture \cite{J.Yu2019} as shown in Fig. \ref{denosing}(b). Different from the traditional autoencoder as shown in Fig. \ref{denosing}(a), we add additive white Gaussian noise to the original signals. The encoder extracts the semantic features of the noise-added signals. The decoder reconstructs the semantic features into the original signals.
The MSE loss is used as the criterion to evaluate the reconstruction result and thus the network has the noise robustness by minimizing this loss. The MSE loss can be expressed as
\begin{align}
\label{Reconstrcution loss}
\mathcal L_{M S E}={\mathbb E}\left({\bm x}_{i}-\hat{\bm{x}}_i^n\right)^{2},
\end{align}
where ${\bm x}_{i}$ denotes the original signal and $\hat{\bm{x}}_i^n$ denotes the reconstructed signal.

\begin{algorithm}[htbp]
	\small
	\caption{Training procedure of the proposed RED method.}
	\label{alg:Training Procedure}
	\textbf {Require}:
	\begin{itemize}
		\item $D$: Training dataset;
		\item $T$: Number of training iterations;
		\item $B$: Number of batches in a training iteration;
		\item $\theta$: Parameters of encoder, classifier and decoder, respectively;
		\item $\theta_{ML}$: The parameter of ML loss;
		\item $lr$: Learning rate of encoder, classifier and decoder, respectively;
		\item $lr_{ML}$: Learning rate of ML loss;
		\item $\lambda_{CE}$, $\lambda_{MSE}$, $\lambda_{ML}$: Scalars for balancing the loss functions;
	\end{itemize}
	
	\begin{algorithmic}
		\Statex {\bf Dataset preprocessing}:
		\State $D \leftarrow \frac{D-\min(D)}{\max(D)-\min(D)}$;
		\For {$t=0$ to $T-1$}:
		\For {$b=0$ to $B-1$}:
		\State Sampling a batch training dataset $({{\bm x}_{i}}, {\bm y_{i}})$ from $D$.
		\Statex \quad\quad\quad{\bf Forward propagation}:
		\Statex \quad\quad\quad{Add artificial noise perturbation to ${\bm x}_{i}$.}
		\State ${\bm x}_i^n=\operatorname{awgn}\left({{\bm x}_{i} };\operatorname{snr}\right)$;
		\Statex \quad\quad\quad{Get the output of encoder, classifier and decoder.}
		\State $\{{\bm z}_i, \hat{\bm y}, \hat{\bm x}_i^n\}= f(\theta^{t, b}, \theta_{ML}^{t,b}; \hat{\bm x}_i^n)$;
		\Statex \quad\quad\quad{Calculate the loss.}
		\State ${\mathcal L}_{CE} = {\mathcal L}_{CE}(\hat{\bm y}_{i}, {\bm y}_{i})$;
		\State ${\mathcal L}_{MSE} = {\mathcal L}_{MSE}({\bm x}_{i},{\hat{\bm x}_i^n})$;
		\State ${\mathcal L}_{ML} = {\mathcal L}_{ML}({\bm z}_{i},{\bm c}_{y_k})$;
		\State ${\mathcal L} = \lambda_{CE}{\mathcal L}_{CE} + \lambda_{MSE}{\mathcal L}_{MSE}+ \lambda_{ML}{\mathcal L}_{ML}$;
		\Statex \quad\quad\quad{\bf Backward propagation}:
		\State $ \theta^{t,b+1} \leftarrow Adam(\nabla_{\theta}, {\mathcal L}, lr, \theta)$
		\State $ \theta_{ML}^{t,b+1} \leftarrow Adam(\nabla_{\theta_{ML}}, {\mathcal L_{ML}}, lr_{ML}, \theta_{ML})$
		\EndFor \State \textbf {end for}
		\EndFor \State \textbf {end for}
	\end{algorithmic}
\end{algorithm}

\subsection{Metric Regularization for High Feature Discrimination}
Feature discrimination is critical for RED. However, the discrimination brought by CE loss is not sufficient, so the ML loss is used as another regularization term. The ML loss \cite{Y.Wen2016} can be formulated as:
\begin{align}
\label{Center loss}
\mathcal L_{M L}=\frac{1}{2}\left\{{\mathbb E}\left\|{\bm {z}}_{i}-\bm{c}_{{y}_{k}}\right\|_{2}^{2}\right\},
\end{align}
where ${\bm z}_{i}$ and ${\bm {c}}_{{y}_{k}}$ denote the semantic feature vectors and the semantic center feature vectors of train samples in the $k$-th category, respectively.
The combination of CE loss and ML loss allows the neural network to extract semantic features with small intra-class distance. The role of ML loss is illustrated in Fig. \ref{Discriminative}.

\subsection{Rogue Emitter Detection}
\begin{figure}[htbp]
	\centering
	\includegraphics[width=3.5 in]{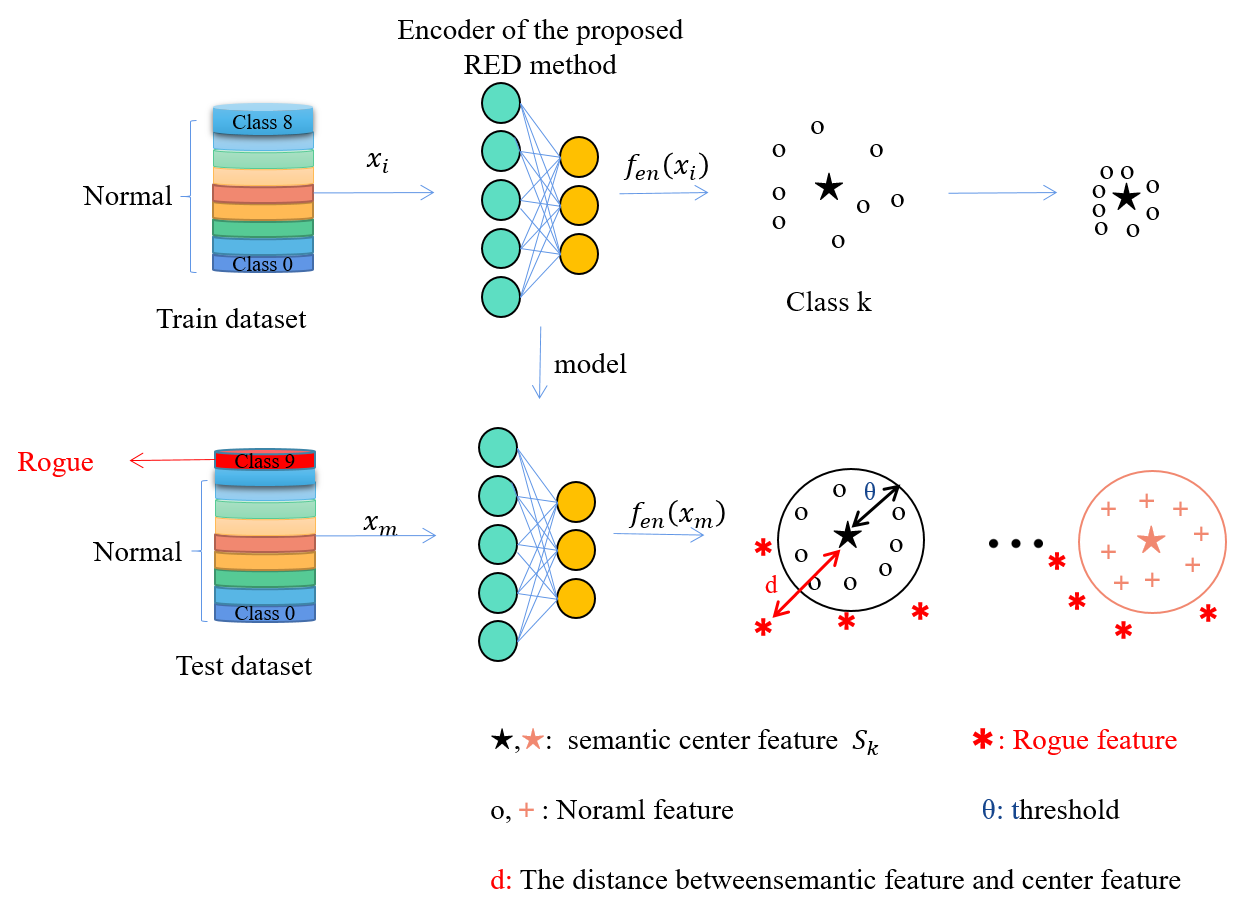}
	\caption{The details of the proposed RED method.}
	\label{Discriminative}
\end{figure}

As mentioned in subsection A, when training is over, only the encoder is retained and used for RED. Fig. \ref{Discriminative} shows a more detailed RED process which can be divided into the following three steps.
\subsubsection{Obtaining semantic center features}
The training dataset is considered as radio signals that emit from the legal emitters and are fed into the encoder to extract the semantic features and the semantic features of the same category are averaged to obtain the semantic center features of each legal emitter. The formula is expressed as ({\ref{centerfeatures}}).

\subsubsection{Calculating the distance between semantic features of radio signals of unknown emitters and the known semantic center features}
Radio signals from unknown emitters are fed into the encoder of the proposed RED method to mine semantic features. The distance between the semantic features of radio signals from unknown emitters and the known semantic center features is calculated, and the formula\cite{Y.Dong2021} is expressed as follows:
\begin{align}
\label{distance}
\begin{array}{l}
{\ d}\left(f_{en}\left({\bm x}_{m}\right), {\bm S}_{k}\right) \\
=\sqrt{\left(f_{e n}\left({\bm x}_{m}\right)-{\bm S}_{k}\right)^{T} {\bm A}_{k}^{-1}\left(f_{e n}\left({\bm x}_{m}\right)-{\bm S}_{k}\right)}
\end{array}
\end{align}
where $f_{en}$ denotes the mapping function of the encoder of the proposed RED method, $f_{en}\left({\bm x}_{m}\right)$ denotes the semantic features of the input signals, and $d$ denotes the Euclidean distance when ${\bm A}_{k}^{-1}$ is the unit matrix.

\subsubsection{Comparing the distance with threshold}
\begin{align}
\label{threshold}
{\left\{\begin{array}{ll}
{\bm x}_{m} \in \text {known}, \!\!\!&\!\!\! \text { if } {{\underset{{\bm S}_{k} \in \bm S}{\min}  {d}\left(f_{en}\left({\bm x}_{m}\right), {{\bm S}_{k}}\right)} \leqslant \lambda \sqrt{{3 t}}} \\
{\bm x}_{m} \in \text {unknown}, \!\!\!&\!\! \!\text { if } {{\underset{{\bm S}_{k} \in \bm S}{\min}  {d}\left(f_{en}\left({\bm x}_{m}\right), {{\bm S}_{k}}\right)}>\lambda \sqrt{{3 t}}}
\end{array}\right.}
\end{align}
where $\lambda$ is the hyperparameter and $t$ is the dimensionality of the semantic center features. If $\min_{{\bm S}_k \in {\bm S}} d(f_{en}({\bm x}_m), {\bm S}_{k})$ is less than or equal to the threshold, the input signal will be judged to belong to the known class, otherwise, the input signal will be judged to belong to the rogue class. The threshold is inspired by the three-sigma rule\cite{F1994}.

\section{Experimental Results}
\subsection{Simulation Parameters}
Our simulations are performed on NVIDIA GeForce GTX1080Ti platform based on pytorch 1.8.1. The maximum epoch $T$ is 150 and the batch size $B$ is 16. We use a dynamic learning rate, set the initial learning rate to 0.001 and change it to one-tenth of the initial rate when the validation loss does not drop for 10 epochs.  Adam is selected as the optimizer. The weighting coefficients $\lambda_{C E}$, $\lambda_{MSE}$, and $\lambda_{ML}$ are set to 1, 0.5 and 0.005, respectively. The threshold $\lambda$ ranges from 0.2 to 0.5 with a step size of 0.05.

\subsection{Dataset Description}
The datasets proposed in \cite{Y.T2021} and \cite{K.Sankhe2019} are used to evaluate the proposed RED method. The dataset in \cite{Y.T2021} is a real radio signal dataset based on a special airborne monitoring system ADS-B. We randomly choose dataset containing 10 classes of aircrafts, where the ratio of the number of known classes to that of rogue classes is 9:1, and the number of samples is 3,736. The dataset in \cite{K.Sankhe2019} is collected from 16 transmitters which are bit-similar USRP X310 radios that emit IEEE 802.11a standards-compliant frames generated via a MATLAB WLAN System toolbox, where the ratio of the number of known classes to that of rogue classes is 15:1, and the number of samples is 53,344. The number of sampling points of both datasets is 6,000, and the format of samples is In-phase/Quadrature (IQ).

\subsection{Evaluation Criteria}
\begin{figure}[htbp]
	\centering
	\includegraphics[width=2.6 in]{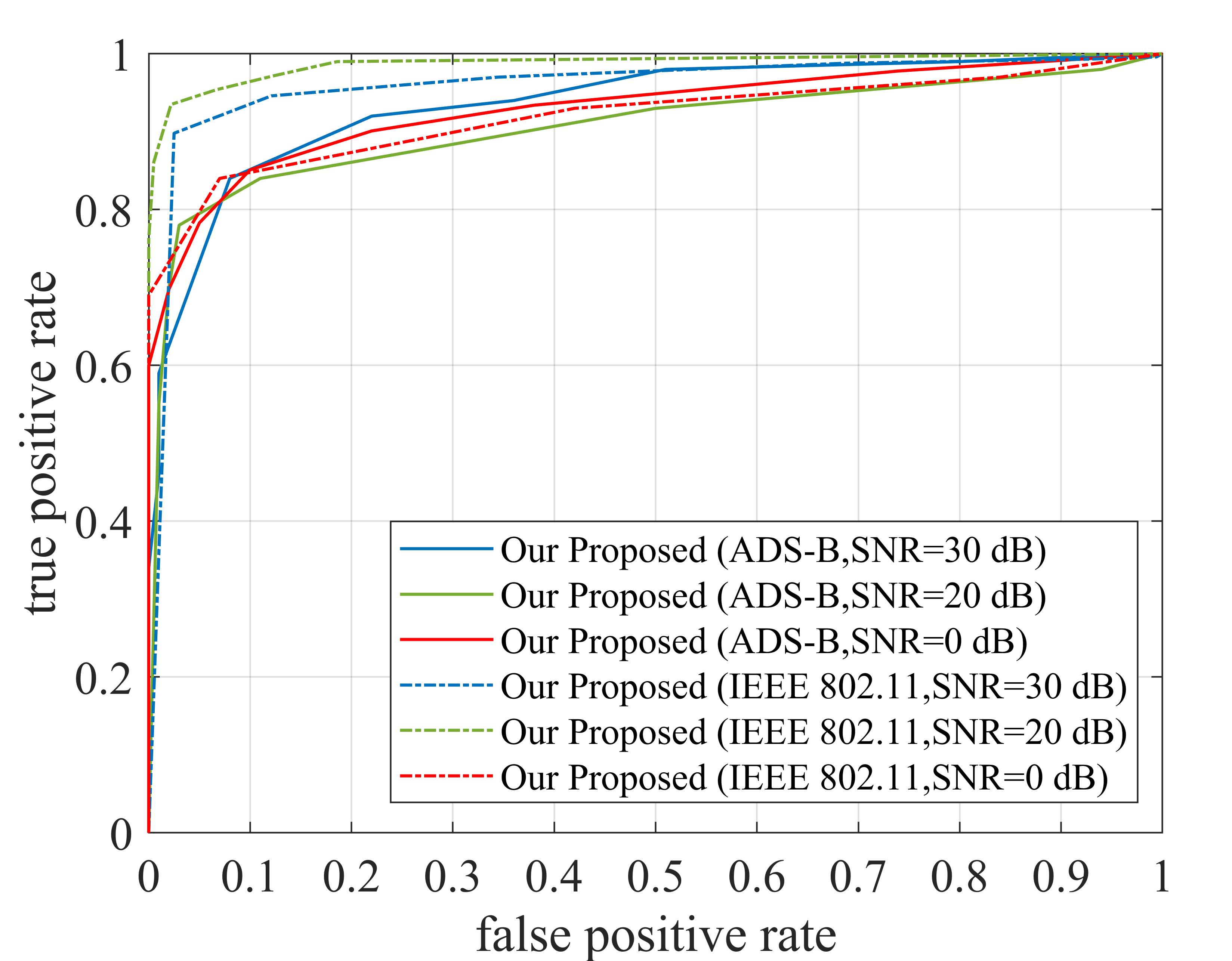}
	\caption{RED performances of the proposed RED method at different SNR.}
	\label{differentSNR}
\end{figure}

\begin{figure}[htbp]
	\centering
	\includegraphics[width=2.6 in]{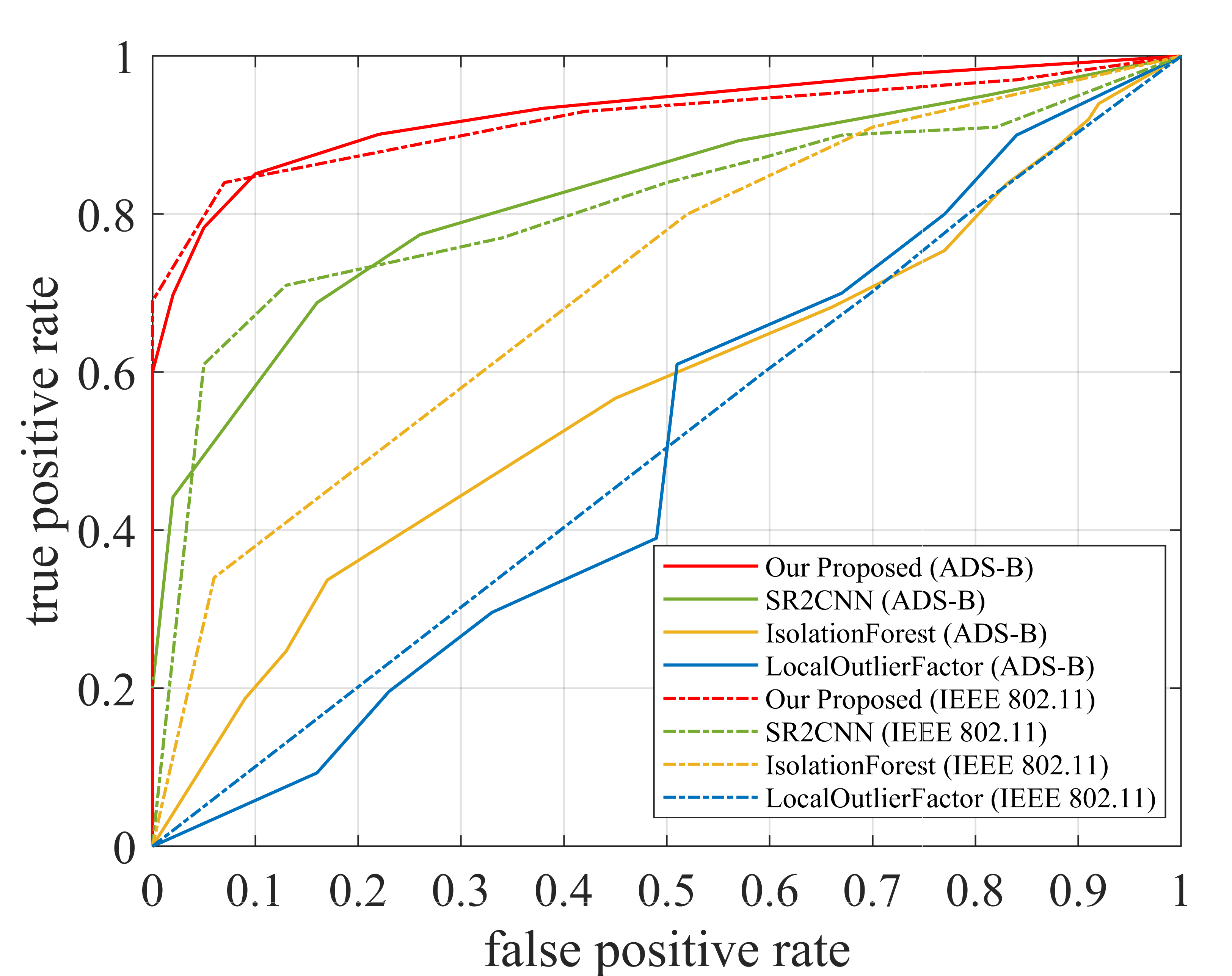}
	\caption{The proposed RED method vs. comparative RED methods when SNR=0 dB.}
	\label{differentmethod}
\end{figure}

\begin{figure}[htbp]
	\centering
	\includegraphics[width=2.6 in]{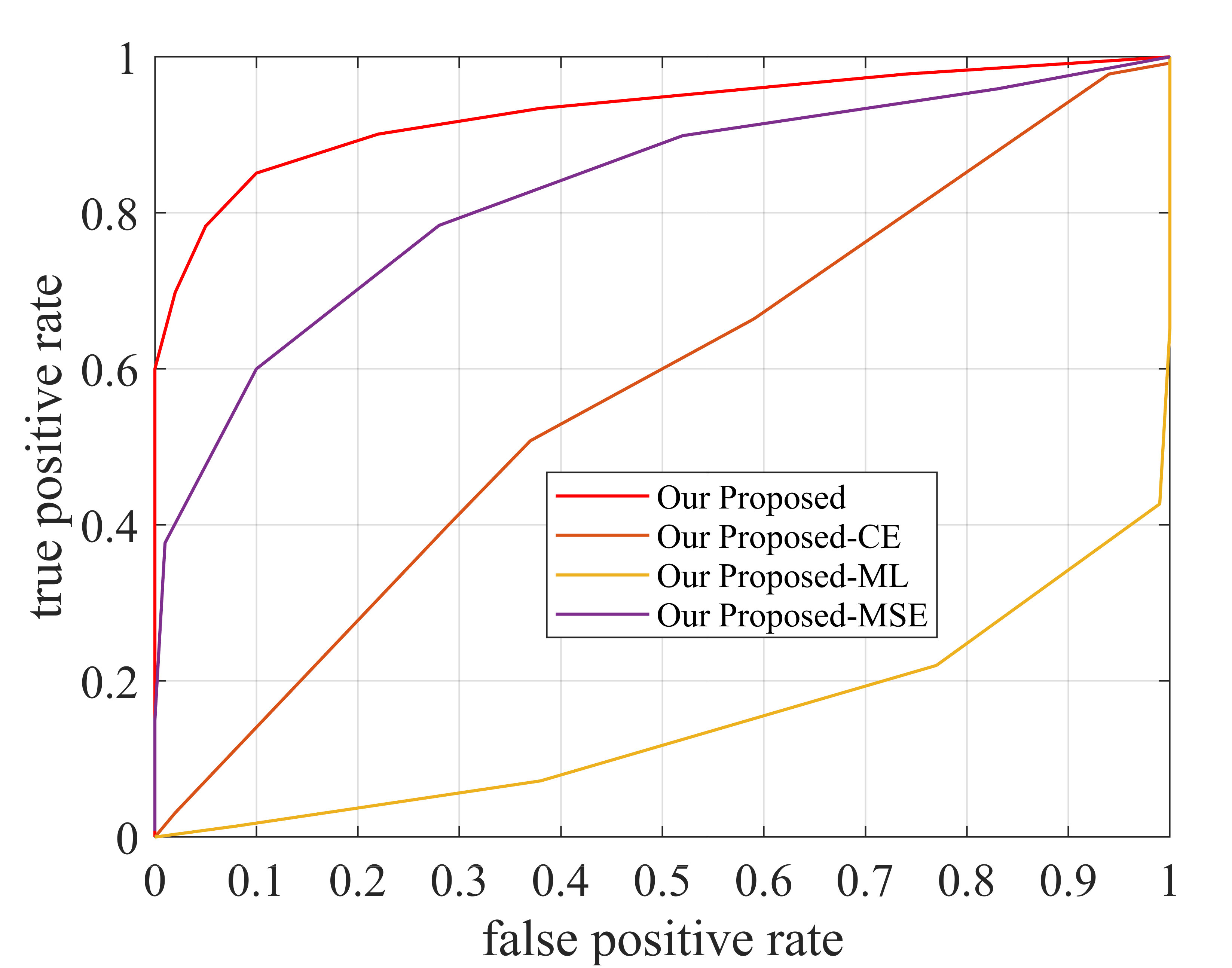}
	\caption{Ablation experiments of proposed method on ADS-B dataset when SNR= 0 dB.}
	\label{Ablationexperiment}
\end{figure}

The receiver operating characteristic (ROC) curve is used to evaluate the performance of RED. The simulation results are plotted in Fig. \ref{differentSNR}, Fig. \ref{differentmethod}, and Fig. \ref{Ablationexperiment} and are analyzed later. The horizontal axis is the false positive rate (FPR) which indicates the probability that the model will determine a rogue device as a known device. The vertical axis indicates the true positive rate (TPR) which indicates the probability that the model will determine a known device as a known device. The mathematical expressions are as follows:
\begin{align}
\label{fun:TPR}
T P R=\frac{T P}{T P+F N}, ~ F P R=\frac{F P}{F P+T N},
\end{align}
where the meaning of each variable is shown in Table. \ref{ROC:variables}. Area under curve (AUC) is used to quantify the ROC, where larger AUC means better RED performance. The mathematical expression of AUC is given as
\begin{align}
	\label{fun: AUC}
	A U C=\int T P R \cdot d(F R P).
\end{align}
Silhouette coefficient (SC) is used to characterize the discrimination of the semantic features.

\begin{table}[htbp]
\caption{The meaning of TPR and FPR variables.}\label{ROC:variables}
\centering
\begin{tabular}{|cc|cc|}
\hline
\multicolumn{2}{|c|}{\multirow{2}{*}{}}         & \multicolumn{2}{c|}{Prediction Value}                                 \\ \cline{3-4}
\multicolumn{2}{|c|}{}                          & \multicolumn{1}{c|}{1}                  & 0                  \\ \hline
\multicolumn{1}{|c|}{\multirow{2}{*}{Real}} & 1 & \multicolumn{1}{c|}{True Positive (TP)}  & False Negative (FN) \\ \cline{2-4}
\multicolumn{1}{|c|}{}                      & 0 & \multicolumn{1}{c|}{False Positive (FP)} & True Negative (TN)  \\ \hline
\end{tabular}
\end{table}

\subsection{Comparative RED Methods}
We compare the proposed RED method with several RED methods, including SR2CNN\cite{Y.Dong2021}, IsolationForest\cite{Liu2008}, and LocalOutlierFactor \cite{Markus2000}.

\subsection{Rogue Emitter Detection Performance}
\subsubsection{RED Performance of the proposed RED method at Different SNR}
To demonstrate the noise robustness of the proposed method, we analyze the detection performance of proposed RED method at SNR $= \{0,20,30\}$ dB. As shown in Fig. \ref{differentSNR} and Table. \ref{AUC:differentSNR}, the ROC curve of the proposed method at SNR $=0$ dB is similar to that at SNR $=30$ dB, which demonstrates that our proposed method has good robustness under different noise.

\begin{table}[htp]
\caption{The AUC of proposed method under different SNR.}\label{AUC:differentSNR}
\centering
\begin{tabular}{|c|c|c|c|}
\hline
SNR/dB      & 0     & 20    & 30     \\ \hline
AUC (ADS-B) & 0.929 & 0.906 & 0.926  \\ \hline
AUC (IEEE 802.11)  & 0.920 & 0.987 & 0.956  \\ \hline
\end{tabular}
\end{table}

\subsubsection{Proposed RED method vs. Comparative RED Methods}
To further demonstrate the noise robustness of proposed RED method, the detection performances of the proposed method and the comparative methods at SNR $=0$ dB are shown in the Fig. \ref{differentmethod} and Table. \ref{AUC:differentmethod}. It is shown that the ROC curve of the proposed method is higher than that of the comparative methods, and the AUC of the proposed method have improved by 0.106, 0.359, and 0.431 on ADS-B dataset and 0.113, 0.207, 0.417 on IEEE 802.11 dataset, respectively, compared to the comparative methods, which shows that the proposed method has better noise robustness than the comparative methods.

\begin{figure}[htbp]
    \centering
    \subfigure[Our proposed]{\includegraphics[width=0.35\hsize]{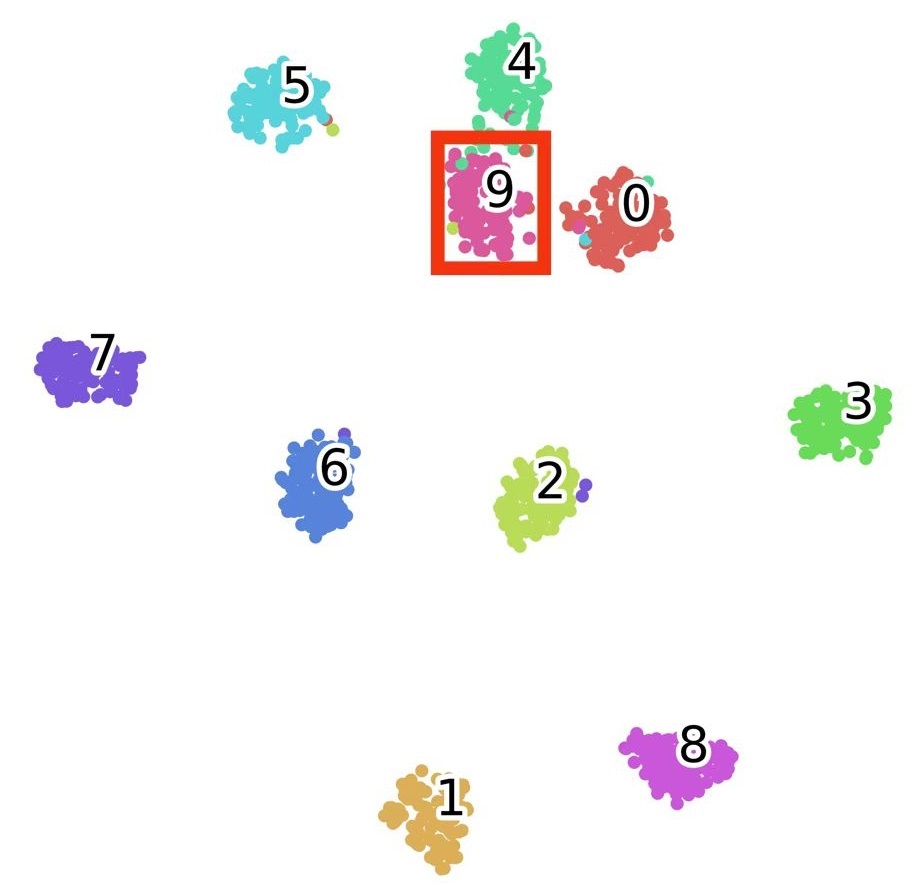}\label{fig: sub_figure1}}\hspace{1cm}
    \subfigure[SR2CNN.]{\includegraphics[width=0.35\hsize]{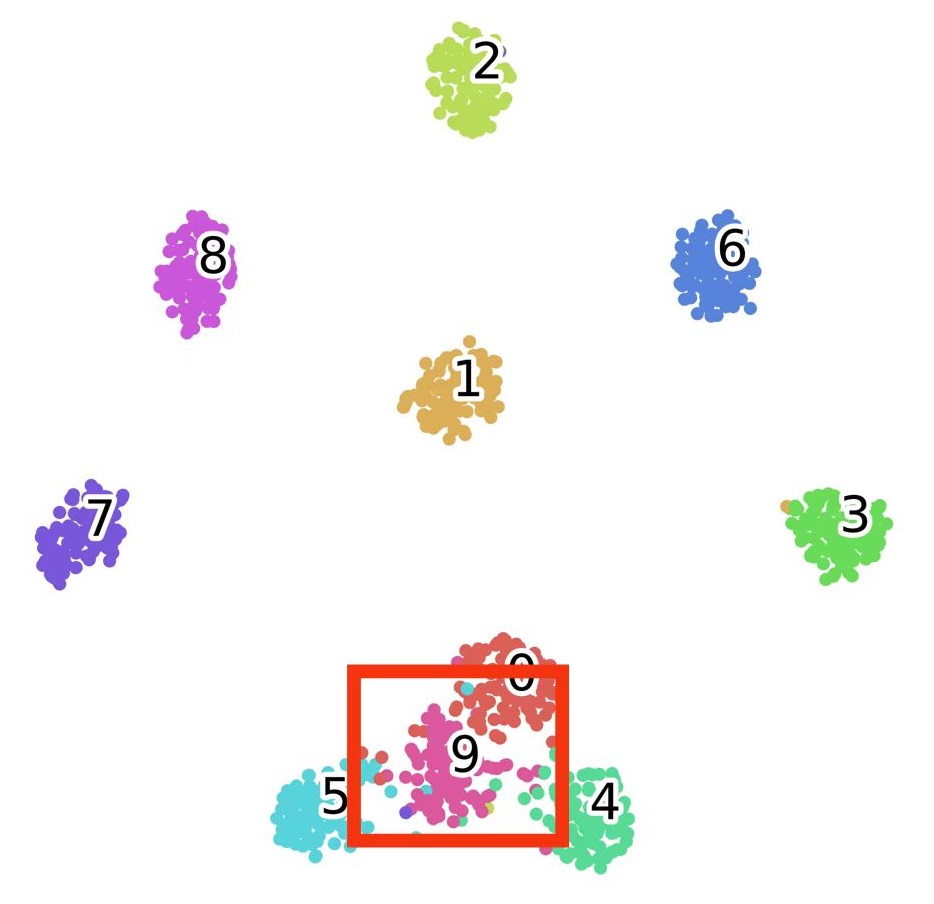}\label{fig: sub_figure2}}
    \caption{The feature visualization of the proposed RED method and SR2CNN testing on ADS-B dataset when SNR $=0$ dB, and the red box indicates the features of rogue category. Note that the SCs of proposed RED method and SR2CNN are $0.4321$ and $0.2955$, respectively.}
    \label{fig: performance of proposed method and SR2CNN}
\end{figure}

To analyze the discrimination of semantic features, t-distributed stochastic neighbor embedding (t-SNE) \cite{L2008} is used to reduce the dimensionality of the extracted semantic features to two dimensions, and the semantic features extracted by the proposed RED method and SR2CNN on ADS-B dataset are compared. As shown by Fig. \ref{fig: performance of proposed method and SR2CNN}, the semantic features extracted by the proposed RED method has higher feature discrimination than that extracted by the SR2CNN at low SNR.

\begin{table}[htbp]
\caption{The AUC of proposed method and other methods.}\label{AUC:differentmethod}
\centering
\begin{tabular}{|c|c|c|c|c|}
\hline
Method      & \textbf{Ours} & SR2CNN & IsolationForest & LOF  \\ \hline
AUC (ADS-B) & \textbf{0.929}          & 0.823  & 0.57            & 0.498                 \\ \hline
AUC (IEEE 802.11)  & \textbf{0.920}          & 0.807  & 0.713           & 0.503                 \\ \hline
\end{tabular}
\end{table}

\subsubsection{Ablation Experiment}
\begin{figure}[htbp]
    \centering
    \subfigure[Our proposed]{\includegraphics[width=0.35\hsize]{DMNet_SNR=0dB_visualization}\label{fig: sub_figure1}}\hspace{1cm}
    \subfigure[Our proposed-CE.]{\includegraphics[width=0.35\hsize]{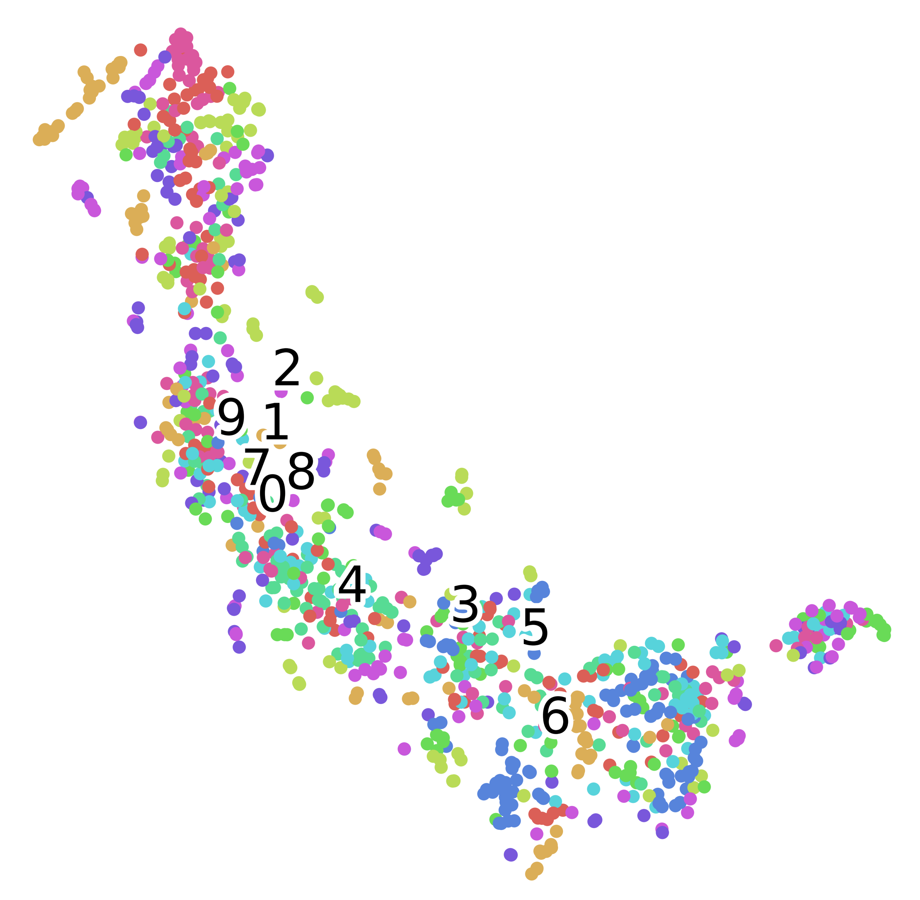}\label{fig: sub_figure2}}
    \vspace{1cm}
    \subfigure[Our proposed-ML.]{\includegraphics[width=0.35\hsize]{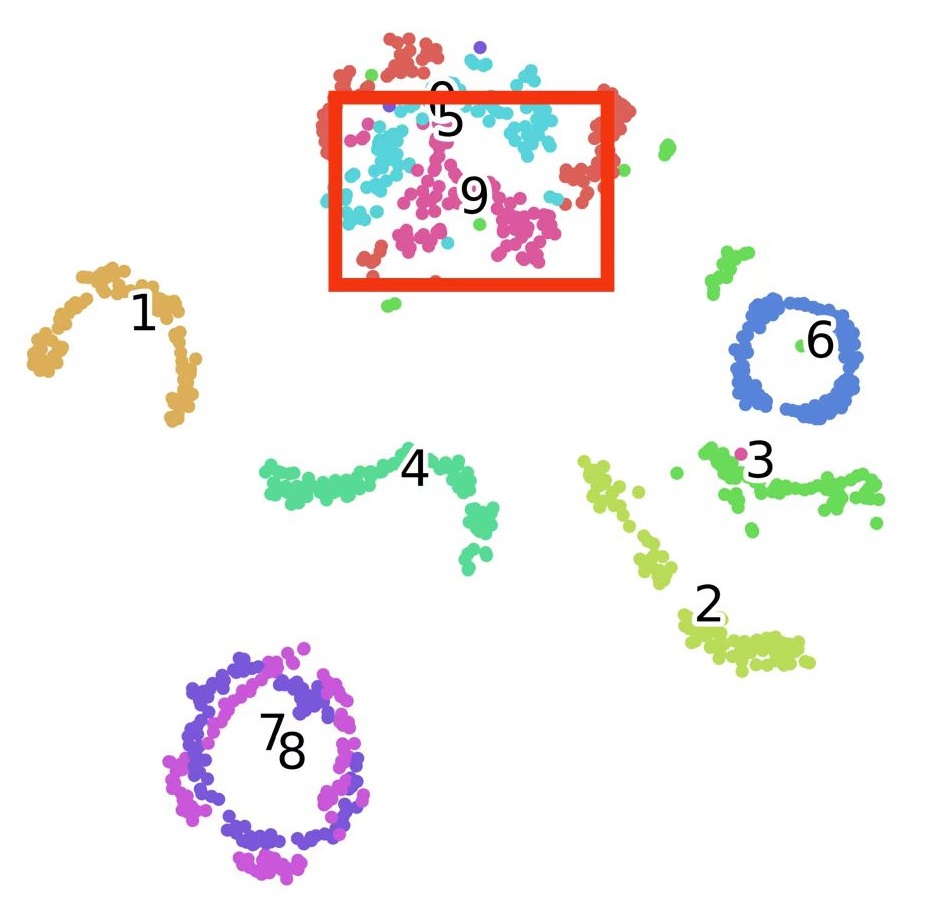}\label{fig: sub_figure3}}\hspace{1cm}
    \subfigure[Our proposed-MSE.]{\includegraphics[width=0.35\hsize]{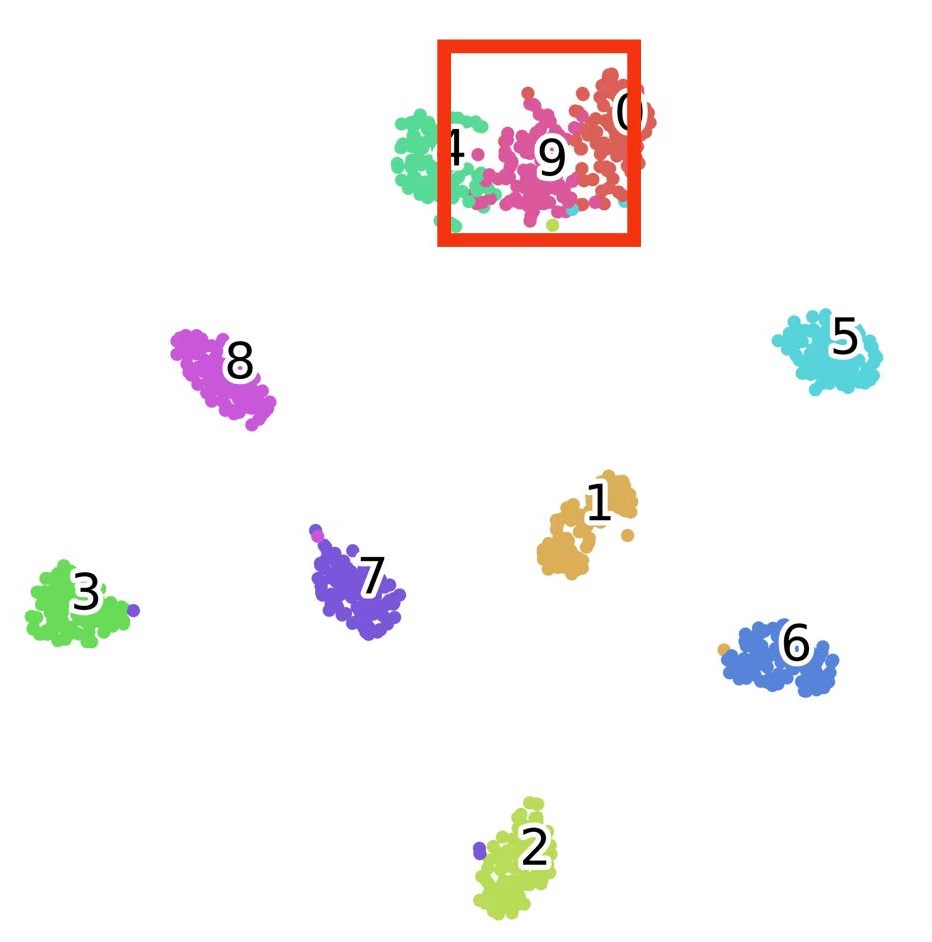}\label{fig: sub_figure4}}
    \caption{The feature visualization of ablation experiments on ADS-B dataset, and the red box indicates the features of the rogue category. Note that the SCs of ablation experiments are $0.4321$, $-0.1979$, $-0.0308$, and $0.4290$, respectively.}
    \label{fig: ablation}
\end{figure}

The objective function of the proposed RED method contains three loss functions, i.e., CE loss, MSE loss, and ML loss. To verify the necessity of each loss, we perform ablation experiments on ADS-B dataset when the SNR is 0 dB. As shown by the Fig. \ref{Ablationexperiment} and Table. \ref{AUC:Ablationexperiment}, without the regularization of the MSE loss, the detection performance of the proposed method decreases, which indicates that MSE loss makes the model have better noise robustness. As shown by the the Fig. \ref{Ablationexperiment} and Fig. \ref{fig: sub_figure1}, \ref{fig: sub_figure3}, without the ML loss, the detection performance of the proposed method decreases, which indicates that ML loss makes the model have better feature discrimination. In addition, through the ablation experiments, we also can observe that MSE loss can not only improve the noise robustness of the model, but also improve the feature discrimination, and ML loss can improve the feature discrimination as well as noise robustness. The proposed method has the best rogue detection performance when all three losses are existing simultaneously.

\begin{table}[htp]
\caption{The AUC of ablation experiments.}\label{AUC:Ablationexperiment}
\centering
\begin{tabular}{|c|c|c|c|c|}
	\hline
	Method      & \textbf{Proposed} & \begin{tabular}[c]{@{}c@{}}Proposed\\ - CE\end{tabular} & \begin{tabular}[c]{@{}c@{}}Proposed\\ - ML\end{tabular} & \begin{tabular}[c]{@{}c@{}}Proposed\\ - MSE\end{tabular} \\ \hline
	AUC (ADS-B) & \textbf{0.929}    & 0.570                                                   & 0.146                                                   & 0.828                                                    \\ \hline
\end{tabular}
\end{table}

\section{Conclusion}
In this paper, we proposed a robust RED method which has strong noise robustness and high feature discrimination. Specifically, a denosing autoencoder is used to reconstruct the noisy signal into the original signal, and an objective function that consists of CE loss, MSE loss, and ML loss is designed to achieve better extraction performance of semantic features with high discrimination while saving feature space. The proposed RED method was evaluated on an open source real-word ADS-B dataset and an IEEE 802.11 dataset and is compared with three RED methods. The simulation results showed that the proposed method has better noise robustness and feature discrimination.

\end{document}